\documentclass[aps,prb,amssymb,reprint]{revtex4-1}

\usepackage{mathtools}
\usepackage[version=3]{mhchem}
\usepackage[english]{babel}
\usepackage[hidelinks,colorlinks=true,linkcolor=blue]{hyperref}
\usepackage[utf8]{inputenc}
\newcommand\at[2]{\left.#1\right|_{#2}}

\draft 

\begin{document}
\title{Ab initio study on lattice thermal conductivity of Cu$_2$O using GGA and hybrid density functional methods}
\author{J. Linnera and A. J. Karttunen}
\email[]{antti.j.karttunen@iki.fi}
\affiliation{Department of Chemistry and Materials Science, Aalto University, P.O. Box 16100, FI-00076 Aalto, Finland}
\date{\today}
	
\begin{abstract}
Lattice thermal conductivity of \ce{Cu2O} was studied using \textit{ab initio} density functional methods. The performance of generalized gradient approximation, GGA-PBE, and PBE0 exchange-correlation functionals was compared for various electronic and phonon-related properties. 3\textit{d} transition metal oxides such as \ce{Cu2O} are known to be a challenging case for pure GGA functionals and in comparison to GGA-PBE the PBE0 hybrid functional clearly improves the description of both electronic and phonon-related properties. The most striking difference is found in the lattice thermal conductivity, where GGA underestimates it as much as 40\% in comparison to experiment, while the difference between the experiment and the PBE0 hybrid functional is only a few percent.
\end{abstract}
	
\maketitle 
	
\section{Introduction}
	
\begin{figure}
	\includegraphics{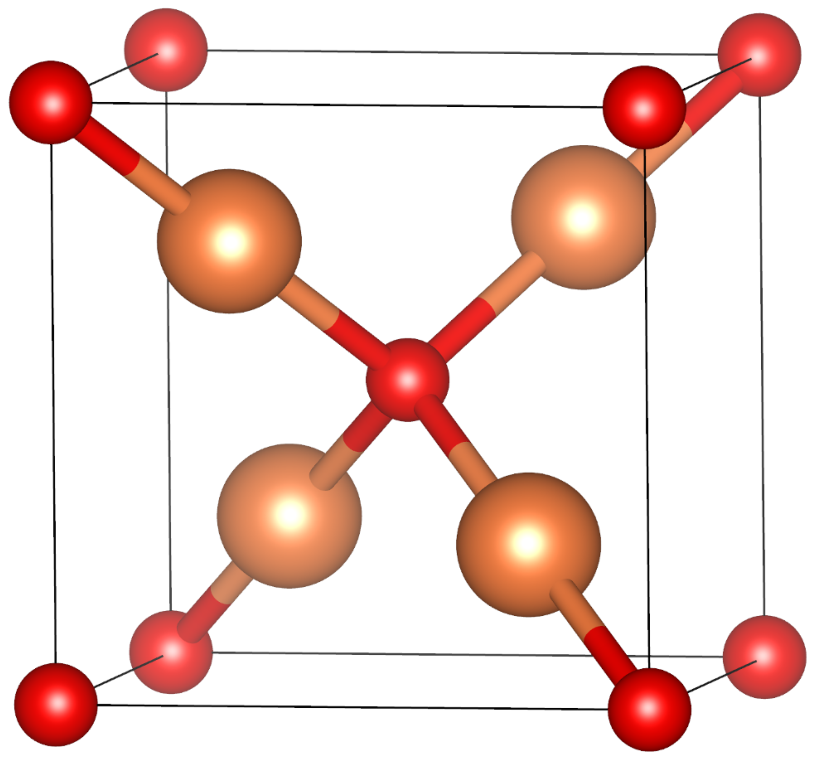}
	\caption{\label{crystal_structure}Unit cell of \ce{Cu2O}. Red: O; brown: Cu.}
\end{figure}

Copper(I) oxide \ce{Cu2O} (Fig. \ref{crystal_structure}) is one of the first semiconductors known to man and has a history of being the first example of different phenomena and devices, such as the first \textit{p}-\textit{n} junction.\cite{Davydov1938,Lashkaryov1941} Copper oxides have been shown to have potential in a plethora of applications such as the \textit{p}-type material in heterojunction photovoltaic cells,\cite{Briskman1992,Minami2011} \ce{CO2} reduction,\cite{Le2011,Frese1991} methanol production catalysis,\cite{Klier} and lately also as a \textit{p}-type thermoelectric (TE) material.\cite{Chen2013,Hartung2015,Mondarte2016} One of the biggest advantages for using copper oxides is the relatively high abundance of Cu and the proficiency of recycling it. For substituting the currently best-performing TE materials such as \ce{PbTe} and \ce{Bi2Te3}, non-toxicity of copper makes it even more attractive for everyday use. In order to use copper oxides as thermoelectric materials, a lot of work is still required to boost the material performance to a useful level.
	
Heat-to-electricity conversion efficiency of a thermoelectric material is defined by a single, dimensionless figure-of-merit $ZT = \sigma S^2 T/\kappa$, where $\sigma$ is the electrical conductivity, \textit{S} is the Seebeck coefficient and $\kappa$ the thermal conductivity of the material. The problem of rationally optimizing materials towards high thermoelectric efficiency is that $\sigma$, \textit{S} and $\kappa$ are very difficult to decouple and systematically improve in a one-at-a-time fashion. One of the most promising techniques has been to decouple $\kappa$ to electronic thermal conductivity $\kappa_e$ and lattice thermal conductivity $\kappa_l$ and then reduce the latter by the means of increasing phonon-phonon scattering via nanostructuring or doping.\cite{Venkatasubramanian1997,Kanatzidis2012} This not only improves \textit{ZT} by shrinking the denominator while keeping electrical properties roughly the same, but can in the best case even enhance them through increased carrier concentrations on appropriate doping levels. Although the procedure is simple on paper, finding the right material compositions and eventually synthesizing them in a controlled fashion is not so straightforward.

With recent advancements in atomistic materials modelling and the ever-increasing computational capacity, it is possible to predict many physical properties within accuracy that is useful for rational materials design. Especially the calculation of lattice thermal conductivities has evolved vastly during the last 10 years from using empirical potentials for silicon to a full \textit{ab initio} treatment of complex materials, several software packages having been developed for this purpose.\cite{Broido2005,Broido2007,Chaput2011,Chaput2013,Li2014,Chernatynskiy2015,Togo2015a,Curtarolo2016,Tadano2014} Albeit these methods and software provide a robust and parameter-free way to gain access to theoretical lattice thermal conductivities, they use density functional theory (DFT) to calculate interatomic force constants, and are thus prone to the common pitfalls of DFT. One of these is the inability of local density approximation (LDA) or generalized gradient approximation (GGA) to correctly describe relatively localized electronic states such as the \textit{d}-states of transition metals.

There are two common methods of handling the troublesome self-interaction error with localized \textit{d}-states: (1) Introducing an on-site Coulomb repulsion to LDA or GGA using Hubbard \textit{U} term, as popularized by Liechtenstein\cite{Liechtenstein1995} and Dudarev\cite{Dudarev1998}, or (2) use a so-called hybrid density functional, where a part of the exchange energy of the system is taken from the exact exchange of Hartree-Fock theory, as introduced by Becke.\cite{Becke1993a,Becke1993b} For 3D periodic systems, the first approach has gained much popularity due to its much lower computational cost when plane-wave basis sets are used. Hybrid functionals are more typically applied in molecular calculations based on local basis functions, where the computational cost for exact exchange is much smaller in comparison to plane waves. However, it is also possibile to apply local basis sets for solid-state systems with periodic boundary conditions, enabling a more cost-effective inclusion of hybrid functionals.

As DFT studies on the lattice thermal conductivity of complex materials become more and more common, it is also interesting to see how DFT-GGA performs in comparison with hybrid functionals in the case of phonon-related properties. We have interfaced the CRYSTAL14 materials modelling package based on Gaussian-type basis sets with Phono3py program package and we report, to our knowledge, the first full hybrid DFT calculations on lattice thermal conductivity.\cite{Dovesi2014,Togo2015a,Tadano2017} The outline of the paper is as follows: First we will discuss the most important aspects of the theory of lattice dynamics related to our calculations. Then we report the results on the basic electronic and structural properties of \ce{Cu2O} followed by phonon-related results such as the phonon dispersion, Grüneisen parameters and lattice thermal conductivity. Finally, we discuss the origins of the performance difference between GGA and hybrid density functional in the case of \ce{Cu2O}.
	
\section{Theoretical background}
\subsection{Lattice dynamics}
Theory of lattice dynamics has been thoroughly discussed in many textbooks.\cite{Ziman,Srivastava,Wallace} Here we outline the theoretical framework for our computational work, introducing the key concepts that are required for the comprehensive analysis of the calculated lattice thermal conductivities. We begin by assuming that crystal potential energy is an analytical function of the atomic displacements from their equilibrium positions. This potential contains all but the kinetic energy of the crystal. Then, with small enough displacements, the potential energy $\Phi$ can be expanded as a power series
	
\begin{equation} \label{potential}
	\Phi = \Phi_0 + \Phi_1  + \Phi_2 + \Phi_3 +...,
\end{equation}
where $\Phi_0$ is the constant potential and other components are as follows:
\begin{align}
	\Phi_1 =& \frac{1}{1!} \sum_{l k \alpha} \Phi_{\alpha}(lk)u_\alpha(lk),\label{p1} \\
	\Phi_2 =& \frac{1}{2!} \sum_{l k \alpha} \sum_{l' k' \beta} \Phi_{\alpha \beta} (lk, l'k') u_\alpha (lk) u_\beta(l'k'),\label{p2} \\
	\label{p3} \Phi_3 =& \frac{1}{3!} \sum_{l k \alpha} \sum_{l' k' \beta} \sum_{l'' k'' \gamma} \Phi_{\alpha \beta \gamma} (lk, l'k', l''k'') \\ \nonumber 
	& \times u_\alpha(lk) u_\beta(l'k') u_\gamma(l''k''),
\end{align}
where \textbf{u}(\textit{lk}) is the displacement of the \textit{k}th atom in the \textit{l}th unit cell from its equilibrium position \textbf{r}(\textit{lk}), \textit{$m_k$} is the atomic mass of atom \textit{k} and Cartesian coordinates are denoted with Greek alphabet. $\Phi_{\alpha \beta}$ and $\Phi_{\alpha \beta \gamma}$ are the second and third order force constants, often called as the harmonic and anharmonic force constants, respectively. They are simply the \textit{n}th derivatives of the potential with respect to the Cartesian coordinates, e.g.
\begin{equation}\label{force_constant}
	\Phi_{\alpha \beta} (lk,l'k') = \at{\frac{\partial^2 \Phi}{\partial u_{\alpha}(lk) \partial u_{\beta}(l'k')}}{0}
\end{equation}
If we would consider only the quadratic term in Eq (\ref{potential}) and the kinetic energy of the system \textit{T}, defined with the displacement derivative as
\begin{equation}\label{kinetic}
	T = \frac{1}{2} \sum_{lk \alpha} m_k [\dot u_\alpha (lk)]^2
\end{equation}
this would constitute the so-called harmonic Hamiltonian \textit{$H_H = \Phi_2 + T$}. In practice the first term $\Phi_1$ is omitted by the assumption that forces on atoms in equilibrium vanish. That is enough to resolve the lattice dynamics from the dynamical matrix eigenvalue problem
\begin{equation} \label{dynamical_eigen}
	\sum_{k' \beta} D_{k k'}^{\alpha \beta} (\textbf{q}) W_{\textbf{q}j}^{\beta k'} = \omega_{\textbf{q}j}^2 W_{\textbf{q}j}^{\alpha k}
\end{equation}
where \textbf{q} is the wave vector, \textit{j} is the band index, $\omega_{\textbf{q}j}$  and $W_{\textbf{q}\textit{j}}$ are the frequency and polarization vector of a phonon mode for a set \{\textbf{q},\textit{j}\}, respectively, and the elements of the dynamical matrix are
\begin{equation} \label{dynamical}
	D_{k k'}^{\alpha \beta} (\textbf{q}) = \sum_{l'} \frac{\Phi_{\alpha \beta}(0k,l'k')}{\sqrt{m_k m_{k'}}} e^{i \textbf{q} \cdot [\textbf{r}(l'k') - \textbf{r}(0k)]}.
\end{equation}
Here and in Eq.(\ref{F}) we have used the translational invariance condition, which states that the force constants depend on \textit{l}, \textit{l}$'$ and \textit{l}$''$ only through their difference, allowing us to drop the sum over \textit{l}.

A more convenient way to write the terms of the potential expansion is to express the displacements with the use of phonon creation and annihilation operators $\hat{a}_{\textbf{q}j}^\dagger$ and $\hat{a}_{\textbf{q}j}$, respectively, resulting in
\begin{equation}\label{displacement}
	\begin{split}
	u_\alpha(lk) =& \left(\frac{\hbar}{2 N m_k}\right)^{\frac{1}{2}} \sum_{\textbf{q}j} \omega_{\textbf{q}j}^{-\frac{1}{2}} (\hat{a}_{\textbf{q}j} + \hat{a}_{-\textbf{q}j}^\dagger) \\
	&\times e^{i \textbf{q} \cdot \textbf{r}(lk)}W_{\textbf{q}j}^{\alpha k}
	\end{split}
\end{equation}
where \textit{N} is the number of unit cells (for full derivation, see e.g. \cite{Srivastava} pp. 93-95). This simplifies the expression for the harmonic Hamiltonian allowing us to write it as a sum of harmonic oscillators
\begin{equation}\label{H_H}
	H_H = \sum_{\textbf{q}j} \hbar \omega_{\textbf{q}j} \left(\frac{1}{2} + \hat{a}_{\textbf{q}j}^\dagger \hat{a}_{\textbf{q}j} \right).
\end{equation}
Furthermore, we obtain the third-order potential $\Phi_3$ from Eq (\ref{p3}) as a sum of three-phonon collisions
\begin{equation}\label{p3_anni}
	\begin{split}
	\Phi_3 =& \sum_{\textbf{q}j} \sum_{\textbf{q}'j'} \sum_{\textbf{q}''j''} F_{\textbf{q}j, \textbf{q}'j', \textbf{q}''j''}^{} (\hat {a}_{\textbf{q}j}^{} + \hat {a}_{-\textbf{q}j}^\dagger) \\ & \times (\hat {a}_{\textbf{q}'j'}^{} + \hat {a}_{-\textbf{q}'j'}^\dagger) (\hat {a}_{\textbf{q}''j''}^{} + \hat {a}_{-\textbf{q}''j''}^\dagger),
	\end{split}
\end{equation}
where the interaction strength is described by the term $F_{\textbf{q}j, \textbf{q}'j',  \textbf{q}''j''}$, written explicitly with phonon modes \{\textbf{q},\textit{j}\} as
\begin{equation}\label{F}
\begin{split}
	&F_{\textbf{q}j, \textbf{q}'j', \textbf{q}''j''} \\
	&=~ \frac{1}{3! \sqrt{N}} \sum_{k k' k''} \sum_{\alpha \beta \gamma} W_{\textbf{q}j}^{\alpha k} W_{\textbf{q}'j'}^{\beta k'} W_{\textbf{q}''j''}^{\gamma k''} \\
	& \times \sqrt{\frac{\hbar}{2m_k \omega_{\textbf{q}j}}} \sqrt{\frac{\hbar}{2m_{k'} \omega_{\textbf{q}'j'}}} \sqrt{\frac{\hbar}{2m_{k''} \omega_{\textbf{q}''j''}}} \\
	& \times \sum_{l'l''} \Phi_{\alpha \beta \gamma} (0k,l'k',l''k'') e^{i \textbf{q}' \cdot [\textbf{r}(l'k') - \textbf{r}(0k)]}\\
	& \times e^{i \textbf{q}'' \cdot [\textbf{r}(l''k'') - \textbf{r}(0k)]} e^{i (\textbf{q} + \textbf{q}' + \textbf{q}'') \cdot \textbf{r}(0k)}\delta_{q + q' + q'',G},
\end{split}
\end{equation}
where $\delta_{q + q' + q'',G}$ is the selection rule for the allowed collisions (e.g. \cite{Wallace} p. 119). In a three-phonon process the wave vectors of the phonons \textbf{q}, \textbf{q}$'$ and \textbf{q}$''$ must add up to a reciprocal lattice vector \textbf{G} in order to preserve crystal momentum (more on the selection rules and their implications, see e.g. \cite{Ziman} pp. 134-145).

\subsection{Thermal conductivity}
Macroscopically thermal conductivity $\kappa$ is defined as the rate of heat energy flow per unit area normal to the thermal gradient $\Delta T$ subjected to a solid material, which together form the heat current
\begin{equation}
	\textbf{Q} = - \kappa \Delta T.
\end{equation}
Generally heat is transferred by several different carriers, and thermal conductivity can be written as a sum of all individual components
\begin{equation} \label{kappa}
	\kappa = \sum_i \kappa_i.
\end{equation}
In solid semiconductor materials it is usually enough to consider heat carried by the phonons ($\kappa_l$) and electrons ($\kappa_e$). Often even the electron contribution can be neglected: with increasing band gap the electrical conductivity decreases, which also decreases the electronic contribution to thermal conductivity according to the Wiedemann-Franz law. In a recent study of \ce{Cu2O} thin films, the electrical conductivity was found to be around 0.5 $\Omega^{-1}$ m$^{-1}$,\cite{Hartung2015} allowing the estimation of the total thermal conductivity to a reasonable accuracy based on the lattice contribution alone. It should be noted that in the case of highly doped semiconductor materials $\kappa_e$ may no longer be negligible.
	
The problem now is to write out a usable formula for describing heat conduction due to phonons. This is done by solving the linearized Boltzmann transport equation (BTE). When the single-mode relaxation time (SMRT) approximation is used, the thermal conductivity for a collection of phonons can be written using their heat capacity \textit{C}, group velocity \textit{v} and their mean free path between collisions $\Lambda$. We can write this for a group of phonons in a finite crystal as
\begin{equation}\label{kinetic_theory}
	\kappa_l = \frac{1}{NV_0} \sum_{\textbf{q}j} C_{\textbf{q}j} v_{\textbf{q}j} \Lambda_{\textbf{q}j},
\end{equation}
where $V_0$ is the volume of the unit cell. Within the harmonic approximation it is already possible to calculate the mode-dependent heat capacity assuming phonons obey Bose-Einstein statistics. Group velocity for the modes obtained from Eq.(\ref{dynamical_eigen}) is just the first derivative of the frequencies with respect to the wave vector. Phonon mean free path can be evaluated as a product of the group velocity and relaxation time of a mode (which is assumed to be the same as the lifetime) $v_{\textbf{q}j} \times \tau_{\textbf{q}j}$, and for the latter we need to include the anharmonicity of the crystal.
	
In the harmonic approximation, phonon lifetimes would be infinite. Anharmonic effects in crystals produce a phonon self-energy $\Delta\omega_{\textbf{q}j} + i\Gamma_{\textbf{q}j}$, where the imaginary part is in fact the reciprocal of the phonon lifetime. An explicit formula for this can be obtained using many-body perturbation theory. Including contributions up to the second-order, the imaginary part can be written as 
\begin{widetext}
	\begin{equation}\label{Gamma}
	\begin{split}
		\Gamma_{\textbf{q}j}(\omega) = \frac{18\pi}{\hbar^2} \sum_{\textbf{q}'j', \textbf{q}''j''} &\left|F_{-\textbf{q}j ,\textbf{q}'j' ,\textbf{q}''j''}\right|^2 \{(n_{\textbf{q}'j'} + n_{\textbf{q}''j''} +1)\delta(\omega - \omega_{\textbf{q}'j'} - \omega_{\textbf{q}''j''})\\
		& +(n_{\textbf{q}'j'} - n_{\textbf{q}''j''})[\delta(\omega + \omega_{\textbf{q}'j'} - \omega_{\textbf{q}''j''}) - \delta(\omega - \omega_{\textbf{q}'j'} + \omega_{\textbf{q}''j''})]\},
	\end{split}
	\end{equation}
\end{widetext}
where terms $n_{\textbf{q}j}$ are phonon occupation numbers.\cite{Maradudin1962,Wallace1966} Phonons are assumed to obey Bose-Einstein statistics and their occupations as a function of temperature are
\begin{equation}\label{occupation}
	n_{\textbf{q}j} = \frac{1}{e^{(\hbar \omega_{\textbf{q}j}/k_B T)} -1}.
\end{equation}

So far everything about phonons and their behavior is general. In actual calculations carried out in this work, the reciprocal space is discretized to a $\Gamma$-point centered mesh of $N_1 \times N_2 \times N_3$ grid points, and the set of considered collisions is thus truncated to events in the vicinity of these grid points. Additionally, when calculating the imaginary part of the self-energy in Eq. (\ref{Gamma}), all other phonon modes are assumed to be in their equilibrium distributions (SMRT approximation, hereafter RTA). We also checked the accuracy of RTA in comparison to full direct solution of the linearized BTE.\cite{Chaput2013} Force constants are calculated using the finite-displacement method, where the dynamical matrix is constructed by calculating forces acting on each atom in a supercell due to the displacement of one atom for the second-order and two atoms for the third-order force constants. Symmetry of the crystal is utilized to reduce the number of required supercell calculations to construct the full set of force constants. Also, only processes of up to three phonons are considered, although the potential energy could be expanded to fourth-order and further terms. The computational cost scales up rapidly already from including third-order interactions, as the number of required force calculations in high-symmetry \ce{Cu2O} is only two for the harmonic part, and 246 for the anharmonic force constants obtained for a 2x2x2 supercell. Anharmonicity is only considered in the calculation of different scattering processes and the effect anharmonicity may have on the phonon eigenvalues is neglected, as well as any higher order force constants and decay pathways involving more than three phonons.

\subsection{Computational details}

All DFT calculations were carried out with CRYSTAL14 program package.\cite{Dovesi2014} Throughout the paper, we compare the performance of two different density functionals: (1) hybrid PBE0 functional with 25\% of Hartree-Fock and 75\% of PBE exchange\cite{Perdew1996,Adamo1999} and (2) PBE GGA functional. In addition, we benchmarked three other well-known functionals to provide a more general picture of the DFT performance for \ce{Cu2O}: (3) hybrid HSE06 functional derived from PBE0 and using an error function screened Coulomb potential for calculating the exchange energy,\cite{Heyd2003} (4)  Minnesota-type meta-GGA functional M06L, and (5) its hybrid counterpart M06 incorporating 27\% exact exchange.\cite{Zhao2006,Zhao2008} For all calculations we used all-electron, Gaussian-type basis sets, mainly triple-$\zeta$-valence + polarization level, based on Karlsruhe def2 basis sets (list of all used basis sets and their derivations are found in the SI).\cite{Weigend2005} Convergence with respect to reciprocal space \textit{k}-sampling was checked and a mesh of $8 \times 8 \times 8$ points was used for all calculations on the primitive unit cell of \ce{Cu2O}.\cite{Monkhorst1976} Coulomb and exchange integral tolerance factors (TOLINTEG) were set to tight values of 8, 8, 8, 8, 16. Default integration grid (XLGRID) was used for the density functional part, along with default total energy convergence thresholds (TOLDEE). TOLDEE was tightened to $10^{-9}$ a.u. for the $\Gamma$-frequency calculation.\cite{Pascale2004,Zicovich-Wilson2004} Elastic constants were calculated with the standard method implemented in CRYSTAL14 within the keyword ELASTCON.\cite{Perger2009,Erba2014} The methodologies used for calculating different physical properties are described in detail in the respective parts of the results section.
	
Harmonic phonon properties and thermal conductivities were obtained using Phonopy and Phono3py program packages, respectively.\cite{Togo2015,Togo2015a} Harmonic force constants were calculated using a $4 \times 4 \times 4$ supercell corresponding to a lattice parameter of 17.27 Å and 384 atoms in the cell, and an atomic displacement of 0.01 Å. For \textit{k}-sampling, only $\Gamma$-point was considered. To confirm that the localized GTO-type basis set provides accurate force constants, we also calculated second-order force constants with Quantum ESPRESSO (QE) plane-wave code both analytically (density-functional perturbation theory, DFPT) and numerically (finite differences with Phonopy). We found good agreement between the GTO and plane-wave results, and detailed results from the QE calculations are included in the Supplemental Material.\cite{SM,QE-2009,GBRV-2014,Gonze1994,DFPT-2001,Kudin2000,Karttunen2015,SCAN2015} When calculating thermal conductivities, the atomic displacement was increased to 0.03 Å for both harmonic and anharmonic force constants. The effect of increasing the atomic displacement was negligible for the harmonic frequencies. A smaller $2 \times 2 \times 2$ supercell with $2 \times 2 \times 2$ \textit{k}-sampling was used for the third-order force constants. \textbf{q}-mesh ($N_1 = N_2 = N_3$) for the calculation of $\kappa_l$ was tested with values varying from 10 to 20 and a mesh of $20 \times 20 \times 20$ \textbf{q} points was used in the final calculations ($18 \times 18 \times 18$ for the full solution of LBTE). A tight $10^{-10}$ a.u. total energy convergence threshold was used in all supercell calculations. All Brillouin zone integrations for calculating phonon density of states (PDOS) and the imaginary part of self-energy have been done with the tetrahedron method. The nonanalytical contribution when \textbf{q} $\rightarrow$ 0 has been taken into account in all phonon calculations.\cite{Wang2010}
	
\section{Results and discussion}
	
\subsection{Structural analysis and band structure}
\ce{Cu2O} is one of the oldest and most studied semiconductor materials, and it crystallizes in the cubic $Pn\bar{3}m$ space group. One unit cell has two formula units and consists of copper atoms linearly coordinated to oxygen atoms, which in turn are tetrahedrally coordinated (Fig \ref{crystal_structure}). Initial structural parameters were taken from a synchrotron radiation study by Kirfel and Eichhorn.\cite{Kirfel1990} The only free structural parameter, lattice constant \textit{a}, was optimized and band structure calculations were performed using three functionals PBE, PBE0 and HSE06 combined with SVP, TZVP and TZVPP level basis sets. Additionally, the mGGA functionals M06L and M06 were tested with the TZVP level basis set and due to similar performance to PBE and PBE0 functionals, respectively, they are discussed in Supplemental Material. As can be seen from Table \ref{cell_parameters}, each combination performs in rather similar fashion in the structural optimization. All approaches result in an increase of \textit{a} that varies between 0.64 -- 1.39 \%. PBE0/TZVP produces a bulk modulus of 107 GPa while the experimental measurements lie in the range of 106 -- 112 GPa.\cite{Manghnani1974,Hallberg1970}
\begin{table}
	\caption{\label{cell_parameters}Optimized lattice parameter \textit{a} (Å) of \ce{Cu2O} for nine different combinations of functionals and Gaussian-type basis sets. The SVP basis set is the smallest, TZVPP the largest. Difference to the experiment (a = 4.269 Å, \textit{T} = 295 K)\cite{Kirfel1990} is shown in parentheses.}
	\begin{tabular}{c | c | c | c}
		&       SVP        &       TZVP       &      TZVPP       \\ \hline
		PBE   & 4.303 (+0.81 \%) & 4.328 (+1.39 \%) & 4.315 (+1.09 \%) \\
		PBE0  & 4.296 (+0.64 \%) & 4.318 (+1.16 \%) & 4.328 (+1.39 \%) \\
		HSE06 & 4.296 (+0.64 \%) & 4.318 (+1.16 \%) & 4.316 (+1.11 \%) \\
	\end{tabular}
\end{table}
	
Functional performance starts to diverge when we take a closer look on the electronic properties. Band structures look in principle identical in all nine cases, but the magnitude of the band gap differs significantly between functionals (Figure S1, Supplemental Material). The difference between the three different basis sets is much smaller. GGAs are somewhat notorious in underestimating band gaps while hybrids more likely tend to overshoot them, a property directly traceable back to the amount of HF exchange used. Band gaps with the TZVP basis set are 0.53 eV, 2.39 eV and 1.87 eV for PBE, PBE0 and HSE06, respectively. The experimental result most often cited is 2.17 eV (varies slightly from 2.0 to 2.2 eV), and the PBE0/TZVP level of theory shows a reasonable agreement with the experiment.\cite{Onsten2007,Baumeister1961} Results and trends obtained here fall in line with previous calculations reported in the literature.\cite{Forte2008,Ruiz1997,Martinez-Ruiz2003,Hu2008,Scanlon2009,Chen2013} 

\begin{figure}
	\includegraphics{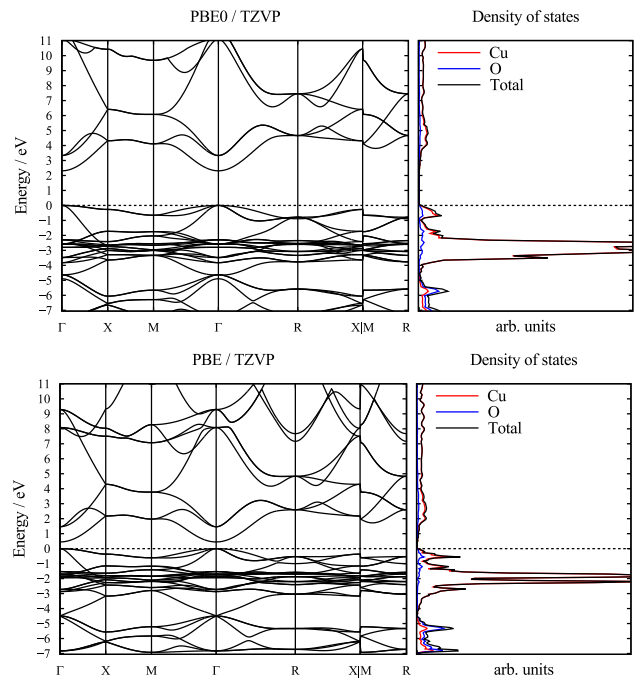}
	\caption{\label{band_dos}Band structure and density of states for \ce{Cu2O} at the PBE0/TZVP and PBE/TZVP level of theory. Dashed line marks the top of the valence bands.}
\end{figure}

The PBE0 and PBE electronic band structures and the atom-projected density of states (DOS) shown in Figure \ref{band_dos} are in line with earlier studies. The valence bands down to -5 eV are dominated by Cu atoms, with some contributions from the O atoms, especially for the topmost valence bands (see Figure S2 in Supplemental Material for corresponding band-projected densities). Below -5 eV, the contribution of the O atoms increases and both Cu and O contribute almost equally. However, comparison of the DOS plots shows that the valence band energies predicted by PBE and PBE0 clearly differ. As discussed before for the PBE functional by e.g. Chen \textit{et al}.,\cite{Chen2013} the incomplete cancellation of the self-interaction of the \textit{d}-orbitals results in them lying too high in energy. The large number of Cu \textit{d}-states between -1 and -4 eV are shifted to almost 1 eV higher in energy with PBE. This shift to higher energies is still seen in the states below -5 eV, although to a smaller degree, because the relative contribution of the oxygen \textit{p}-orbitals is larger. As discussed in the following section, the differences in the PBE0 and PBE electronic structure also result in clear differences in the lattice dynamical properties of \ce{Cu2O}.

\subsection{Lattice dynamics and phonon dispersions}

Harmonic frequencies at the $\Gamma$-point vary a lot depending on the used method of calculation. Table \ref{gamma_frequencies} extends the list of vibrational frequencies of \ce{Cu2O} found in Ref \cite{Sanson2011} with the computational results obtained in this study from Phonopy. $\Gamma$-frequencies were calculated also with the built-in phonon routines of CRYSTAL14, and since the results were practically identical with the results from Phonopy, they are not shown here. Vibrational energies from PBE0 are clearly closer to the experiment in comparison to PBE. A noteworthy detail is the systematicity of this study and results from Ref \cite{Bohnen2009} in column (d). In these studies all values are either within the range of experimental measurements or below them, whereas in other studies some modes are found lower and some higher in energy than the experiment.

\begin{table}
	\caption{\label{gamma_frequencies}Frequencies of phonon modes at $\Gamma$-point. PBE and PBE0 mark the frequencies obtained in this study with given functionals, (a) - (d) are other computational results and column (e) lists experimental values. All values are in wavenumbers (cm$^{-1}$).}
	\begin{tabular}{l | c  c  c  c  c  c  c}
		Mode               &  PBE  &  PBE0 &    (a)\cite{Sanson2011}  &    (b)\cite{Carabatos1971}  &   (c)\cite{Mittal2007} &   (d)\cite{Bohnen2009}  &  (e)\cite{Petroff1975,Ivanda1997} \\ \hline
		$F_{2u}$           &   64  &   86  &   67  &   99  &  101  &   72  & 86-88   \\
		$E_u$              &   79  &  108  &  119  &  100  &  150  &   86  & 109-110 \\
		$F_{1u}$ (1) (TO)  &  133  &  144  &  142  &  143  &  115  &  147  & 146-153 \\
		$F_{1u}$ (1) (LO)  &  135  &  148  &  146  &  159  &  145  &  148  & 149-154 \\
		$B_u$              &  334  &  340  &  350  &  307  &  328  &  337  & 350     \\
		$F_{2g}$           &  491  &  497  &  515  &  549  &  515  &  496  & 515     \\
		$F_{1u}$ (2) (TO)  &  603  &  619  &  635  &  608  &  578  &  609  & 609-640 \\
		$F_{1u}$ (2) (LO)  &  627  &  655  &  654  &  639  &  617  &  629  & 638-665 \\
	\end{tabular}
\end{table}

\begin{figure}
	\includegraphics{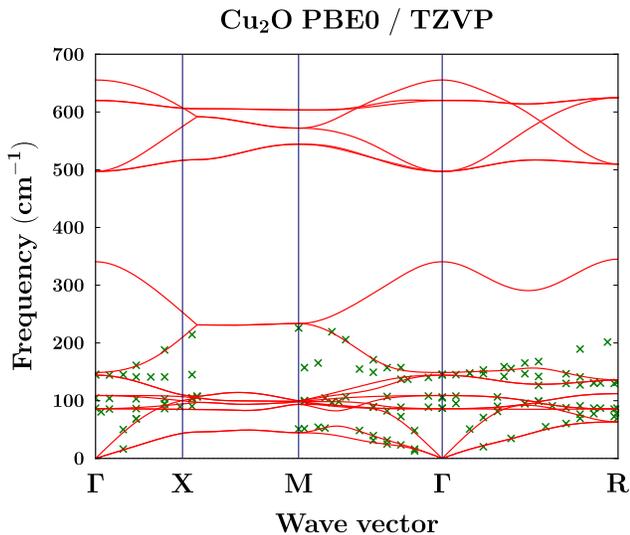}
	\caption{\label{beg_dispersion}Phonon dispersion of \ce{Cu2O} for the relevant paths in the reciprocal space (PBE0/TZVP level of theory). Crosses are data points from Ref \cite{Beg1976} and lines are theoretical results obtained in this study.}
\end{figure}

Full phonon dispersion within the harmonic approximation was calculated for \ce{Cu2O} with PBE, PBE0 and M06 for their optimized unit cells. Due to the similarity of PBE0 and M06 results, the latter are discussed only in the Supplemental Material (Figure S4). Considering first the dispersion obtained using the hybrid PBE0 functional, it is in excellent agreement with the early inelastic neutron scattering experiments by Beg and Shapiro, apart from a few data points.\cite{Beg1976} One phonon band in each of the measured paths around 150 cm$^{-1}$ differs markedly from the otherwise near perfect match as seen in Fig \ref{beg_dispersion}. These same discrepancies were discovered by Bohnen \textit{et al}. in their calculations and they deduced it to be an error in the earlier experiments.\cite{Bohnen2009} To verify this, they carried out a new set of neutron scattering measurements which indeed proved to be different along those reciprocal space paths, and support both their and our theoretical results. Non-analytic correction leads to LO-TO splitting in the IR active modes at 144 and 620 cm$^{-1}$ with values of 4.5 and 35.5 cm$^{-1}$, respectively. These are somewhat larger than the experimental values of 3 and 29 cm$^{-1}$. PBE on the other hand produces smaller values, 2.7 and 23.9 cm$^{-1}$, similar to previous PBE calculations.\cite{Bohnen2009,Dawson1973}
	
\begin{figure*}
	\includegraphics{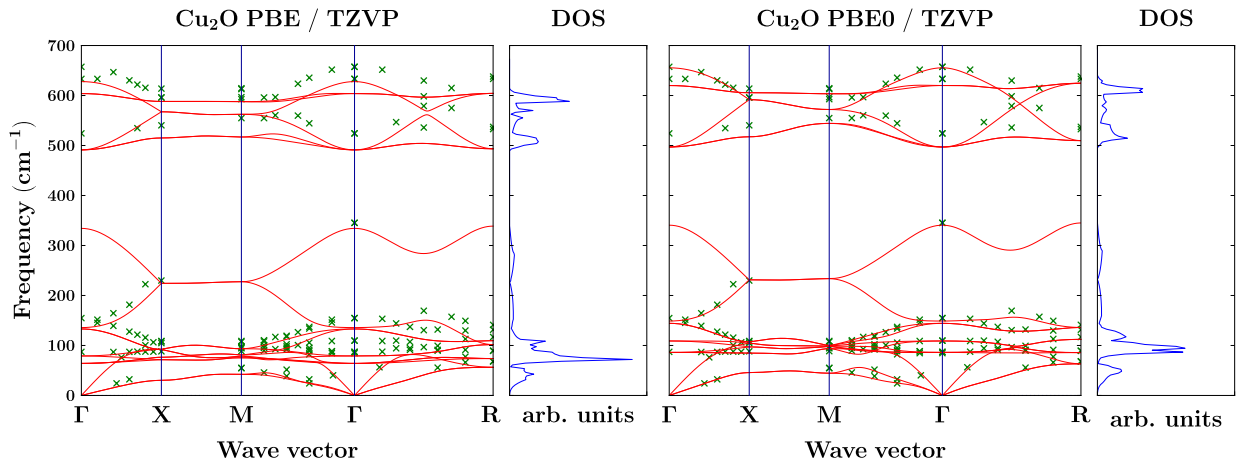}
	\caption{\label{bohnen_dispersion}Phonon dispersion for the relevant paths in the reciprocal space. Crosses are data points from Ref \cite{Bohnen2009} and lines are theoretical results obtained in this study. PBE/TZVP results are shown on the left and PBE0/TZVP results on the right. While the DOS is plotted in arbitrary units, the x-axes are normalized between the DOS plots to show the actual differences in DOS.}
\end{figure*}

When the phonon dispersion throughout the Brillouin zone is considered, a similar softening of modes in comparison to experiment can be seen for PBE. Phonon dispersions and phonon density of states (PDOS) for both PBE and PBE0 functionals are plotted together with the experimental data from Ref \cite{Bohnen2009} in Figure \ref{bohnen_dispersion}. Even though bands from PBE calculations are in some cases closer to experimental points than PBE0, this can probably be attributed to fortuitous cancellation of errors. Generally, the band shapes are closely alike with both functionals, although some tendency can be seen with PBE to flatten the bands as opposed to the higher dispersion obtained with PBE0. This is immediately reflected upon the group velocities of the harmonic phonons, e.g. for transverse acoustic modes right near the $\Gamma$-point, group velocities obtained with PBE0 are around 20\% larger than those obtained with PBE. The PDOS is clearly affected by the softening of the low-frequency modes with PBE. The highest PDOS is found about 20 cm$^{-1}$ lower with the PBE functional in comparison to PBE0 and the shape of the PDOS feature is clearly different for the two functionals.
	
\subsection{Lattice thermal conductivity}
\begin{figure}
	\includegraphics{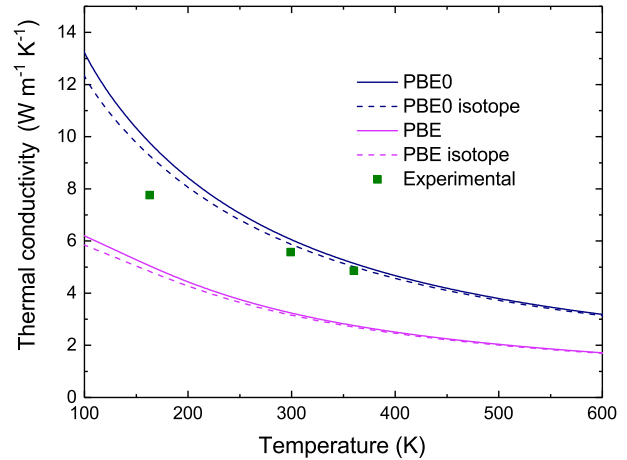}
	\caption{\label{thermal_conductivity}Lattice thermal conductivity $\kappa_l$ of \ce{Cu2O} between 100 and 600 K. Computational results have been obtained at the PBE/TZVP and PBE0/TZVP levels of theory, while the experimental values are from Ref. \cite{CRC}.}
\end{figure}

Moving from phonon dispersions to lattice thermal conductivities results in even larger differences between the GGA-PBE and hybrid PBE0 functionals. Unfortunately, there are only few experimental single crystal data available for the lattice thermal conductivity of \ce{Cu2O}. However, Figure \ref{thermal_conductivity} shows a clear difference between PBE and PBE0, something that could be expected from the differing phonon dispersion results. PBE clearly underestimates the lattice thermal conductivity of \ce{Cu2O} as the calculated values at 300 and 360 K are 3.2 and 2.8 W m$^{-1}$ K$^{-1}$, respectively, more than 40\% below the experimental values of 5.6 and 4.9 W m$^{-1}$ K$^{-1}$.\cite{CRC} PBE0, on the other hand, overshoots the experimental values only about 7\% with predictions of 6.0 and 5.1 W m$^{-1}$ K$^{-1}$. Since \ce{Cu2O} is cubic, thermal conductivity is the same in all directions. Full solution of the LBTE gives very similar results as the RTA. At 300 K, $\kappa_l$ with PBE full solution is the same 3.2 W m$^{-1}$ K$^{-1}$ as with RTA and with PBE0 $\kappa_l$ is increased from 6.0 to 6.1 W m$^{-1}$ K$^{-1}$. Dashed lines in Figure \ref{thermal_conductivity} show the lattice thermal conductivities when isotope scattering is included in the calculations. This introduces a mass variance term based on natural isotope distributions in the respective elements (for details, see Ref. \cite{Togo2015a}). As a result, $\kappa_l$ decreases slightly, around a few percents with PBE0 and a little less with PBE, bringing the predictions closer to experimental values with the hybrid PBE0 functional and further away from experiment with the GGA-PBE. The predicted lattice thermal conductivity at 300 K for PBE0 with isotope scattering is 5.9 W m$^{-1}$ K$^{-1}$, and for PBE it rounds up to the same value of 3.2 W m$^{-1}$ K$^{-1}$ as without the isotope scattering.
	
Comparison of the lattice thermal conductivities with the experimental results at 160 K seems to bring PBE closer to the experiment and PBE0 away from it. This isn't, however, necessarily a result of one functional performing better than the other. When moving to lower temperatures, defect scattering due to point defects and other defects such as dislocations starts to play an increasingly important role in overall thermal resistance, while in higher \textit{T} phonon-phonon scattering dominates. Boundary scattering can also play a role in lower temperatures, but our calculations show that in the case of \ce{Cu2O} the grains would need to be clearly smaller than one micrometre to explain the difference at 160 K (the experimental data are for a single crystal). Full solution of the LBTE at 160 K does not improve on the RTA result, showing only 2\% increase in $\kappa_l$. \ce{Cu2O} is rarely stoichiometric, so the computational results shown in Figure \ref{thermal_conductivity} lack a scattering mechanism that could further decrease the lattice thermal conductivity in lower temperatures. Non-stoichiometricity of \ce{Cu2O} has been studied theoretically in quite detail, and the feature-rich Raman spectra obtained regardless of synthesis methods gives a strong indication of naturally occurring defects (only one vibrational mode is Raman active in pure \ce{Cu2O}).\cite{Wright2002,Nolan2006,Soon2009,Sander2014}

As we calculate thermal conductivity through Eq. \ref{kinetic_theory}, it is possible to study the individual factors giving rise to the differences between the PBE and PBE0 functionals. Heat capacities for \ce{Cu2O} are virtually identical with both methods (Fig S7, Supplemental Material), suggesting that the disparity must be attributed to differences in group velocities and phonon lifetimes. Except for the third-order force constants $\Phi_{\alpha \beta \gamma}$ in Eq. \ref{F}, both group velocities and phonon lifetimes are calculated from the harmonic phonon eigenfrequencies and eigenvectors with no anharmonic corrections to phonon eigenfrequencies. Thus, it is very important to have accurate phonon dispersion relations and PDOS as a basis for the calculation of phonon-phonon interactions. In addition to (1) the accuracy of the DFT forces, it is also important to keep in mind other possible sources of errors in comparison to the experiment: (2) point defect scattering due to impurities or non-stoichiometry is neglected and (3) the collision space for determining the processes producing thermal resistance is described using a finite \textbf{q}-mesh. The total contribution arising from these three sources is not easy to estimate and it is possible that some cancellation of errors is also present.
	
\begin{figure}
	\includegraphics{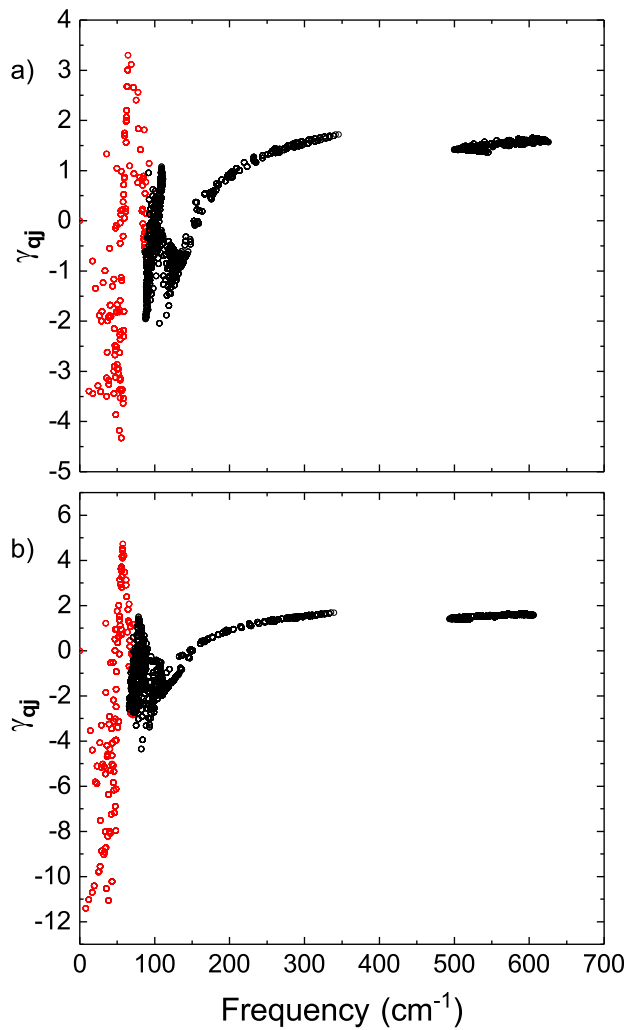}
	\caption{\label{gruneisen}Mode-Grüneisen parameters calculated with a) PBE0 and b) PBE functionals, plotted against frequency. Red circles mark the acoustic modes and black circles are optical modes. Notice the different y-axes for PBE and PBE0 functionals.}
\end{figure}

For a symmetric material with a simple structure, \ce{Cu2O} has a rather low lattice thermal conductivity. This is usually consistent with compounds that have large mode-dependent Grüneisen parameters, a measure for the anharmonicity of phonons in the system.\cite{Heremans2008,Wolverton2012} In fact, in continuum theory the phonon lifetime $\tau_{\textbf{q}j}$ is inversely proportional to the square of the averaged Grüneisen parameter $\gamma$.\cite{Ziman} The dimensionless mode-Grüneisen parameter is often written by relating the shifts in phonon frequencies of modes \{\textbf{q},j\} with respect to varying volume, but a more general way is to define it using strain $\eta_{\mu \nu}$
\begin{equation}
	\gamma_{\mu \nu}(\textbf{q}j) = - \frac{1}{\omega_{\textbf{q}j}} \frac{\partial \omega_{\textbf{q}j}}{\partial \eta_{\mu \nu}}.
\end{equation}
Within our DFT-based approach, $\gamma_{\textbf{q}j}$ can be obtained from third-order force constants using the relation [Ref. \cite{Wallace} pp. 204-205]
\begin{equation}\label{mode-gruneisen}
\begin{split}
	\gamma_{\textbf{q}j} =& - \frac{1}{2 \omega_{\textbf{q}j}^2} \sum_{l' l''} \sum_{k k' k''} \sum_{\alpha\beta\gamma} \frac{W_{\textbf{q}j}^{\alpha k \ast} W_{\textbf{q}j}^{\beta k'}}{\sqrt{m_k m_{k'}}}\\
	& \times \Phi_{\alpha \beta \gamma}(0k, l' k', l'' k'') \\
	& \times e^{i\textbf{q} \cdot [\textbf{r}(l' k') - \textbf{r}(0 k)]} \times\textbf{r}(l'' k'' \gamma).
\end{split}
\end{equation}
Calculated mode-Grüneisen parameters are shown in Figure \ref{gruneisen}. Data shown in these plots have been obtained using a $10 \times 10 \times 10$ \textbf{q}-mesh. Convergence check with respect to a $16 \times 16 \times 16$ mesh showed changes only in the density of data points. PBE0 results are in agreement with previous \textit{ab initio} results obtained using quasi-harmonic approximation (QHA). The only notable difference is in the maxima of the modes around 60 cm$^{-1}$, where our calculations show values of 3.5 and previous QHA results reach values as high as 4.5. The $\gamma_{\textbf{q}j}$ obtained with PBE0 also compare well with the experimental values of Reimann \textit{et al}. determined from Raman measurements, which range from -3.4 to +1.7.\cite{Reimann1989} With PBE, the maximum value is around 5 and minimum is -11.5, which is unreasonably low. Distribution of $\gamma_{\textbf{q}j}$ with respect to frequencies becomes almost identical for $\omega \geq$ 200 cm$^{-1}$. We would like to note that there is no way to distinguish different modes when the phonon bands cross, and thus the division into acoustic and optical modes in Figure \ref{gruneisen} is based on the listing provided by Phono3py.

Based on equation(\ref{mode-gruneisen}), according to which $\gamma_{\textbf{q}j}$ are calculated, the reason for the observed discrepancies between PBE and PBE0 is two-fold. Smaller harmonic frequencies of PBE show up as larger mode-Grüneisen parameters since $\gamma_{\textbf{q}j} \propto \omega_{\textbf{q}j}^{-2}$. This alone might not explain the difference of over a factor of two, although the relative difference in the lower end of harmonic frequencies between functionals PBE and PBE0 is greater than in higher wavenumbers. This results in greater differences in absolute values of $\gamma_{\textbf{q}j}$. As the vibrational modes associated with the phonons are the same, meaning $e^{i\textbf{q}\cdot \textbf{r}}$ and polarization vectors cannot differ significantly, the rest of the differences can be attributed to differences in third-order force constants $\Phi_{\alpha \beta \gamma}$.

\begin{figure}
		\includegraphics{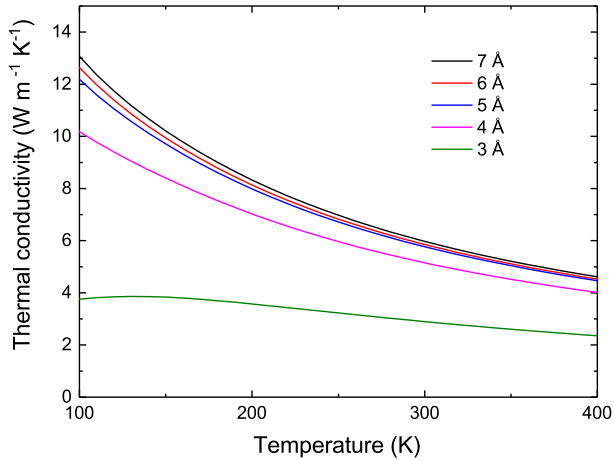}
		\caption{\label{cutoff}Lattice thermal conductivity calculated with PBE0 using different distance cutoff values for the third-order force constants (normally, all third-order force constants within the supercell were calculated).}
\end{figure}

To study the range of phonon-phonon interactions in \ce{Cu2O}, we also calculated the lattice thermal conductivity with a set of cutoffs, setting the respective third-order force constants to zero if the separation of any atom pair in a triplet of atoms is longer than the cutoff value. Figure \ref{cutoff} shows the calculated $\kappa_l$ at temperatures between 100 and 400 K with different cutoff values using PBE0. It is immediately apparent that a cutoff distance of 3 Å will not suffice, as it will only take into account forces between the nearest neighbor atoms. Majority of the strongest interactions are captured already with a 4 Å cutoff, including next-nearest neighboring atom pairs and oxygens in adjacent oxygen tetrahedra. For accuracy's sake, however, a cutoff distance of at least 5 Å is needed, beyond that the effects are negligible. If knowledge on the range of interactions would be available prior to forming the supercells with finite displacements, the number of required force calculations could be drastically reduced. That said, setting cutoff ranges without testing should generally be avoided. Although some estimates can be made from the structure by choosing a number of neighbor shells to include, a good choice is always heavily material-dependent, e.g. a cutoff of 3.8 Å that is sufficient for \ce{WSe2} would not capture all the relevant interactions in \ce{Cu2O}.\cite{Lindroth2016}

\begin{figure}
	\includegraphics{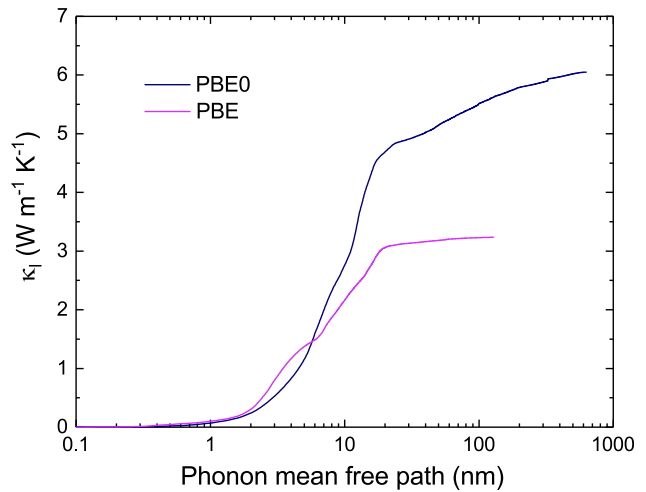}
	\caption{\label{cumulative}Cumulative $\kappa_l$ with respect to phonon mean free path $\Lambda_{\textbf{q}j}$ calculated at 300 K with functionals PBE and PBE0.}
\end{figure}

We also calculated the cumulative lattice thermal conductivity $\kappa_{\textbf{q}j}(\Lambda)$, where only modes with a mean free path smaller than a certain threshold contribute to the thermal conductivity. In Figure \ref{cumulative} we have plotted the cumulative lattice thermal conductivity with respect to phonon mean free path $\Lambda_{\textbf{q}j}$ for both PBE and PBE0. As the absolute value of $\kappa_l$ is smaller for PBE, it is natural for it to reach its maximum at much shorter mean free paths compared to PBE0. Shapes of the curves are rather similar for PBE0 and PBE, with some obvious numerical differences. The curve resembles those of other semiconductors, such as \ce{PbTe}\cite{Lindroth2016} or ZnO \cite{Karttunen2016} with a slow start before the mean free path grows to relevant magnitudes, followed by a somewhat steep linear regime and finally a less steep plateau towards the maximum. For both \ce{Cu2O} and \ce{PbTe} the majority of heat is carried by phonons with a mean free path less than 20 nm. For ZnO the linear regime continues through the plot, indicating a more even distribution of heat carrying phonons over the whole frequency range. The slope of the plot can be used as an approximate measure of where the density of heat carrying phonons lies with respect to $\Lambda_{\textbf{q}j}$. Figure \ref{cumulative} does not explicitly show the density of modes, but when comparing two curves for the same material, general features, such as the bulge with PBE at $\Lambda_{\textbf{q}j}$ $\leq$ 5 nm, reveal that the number of modes associated with such mean free paths is greater with PBE than with PBE0.

\begin{figure}
	\includegraphics{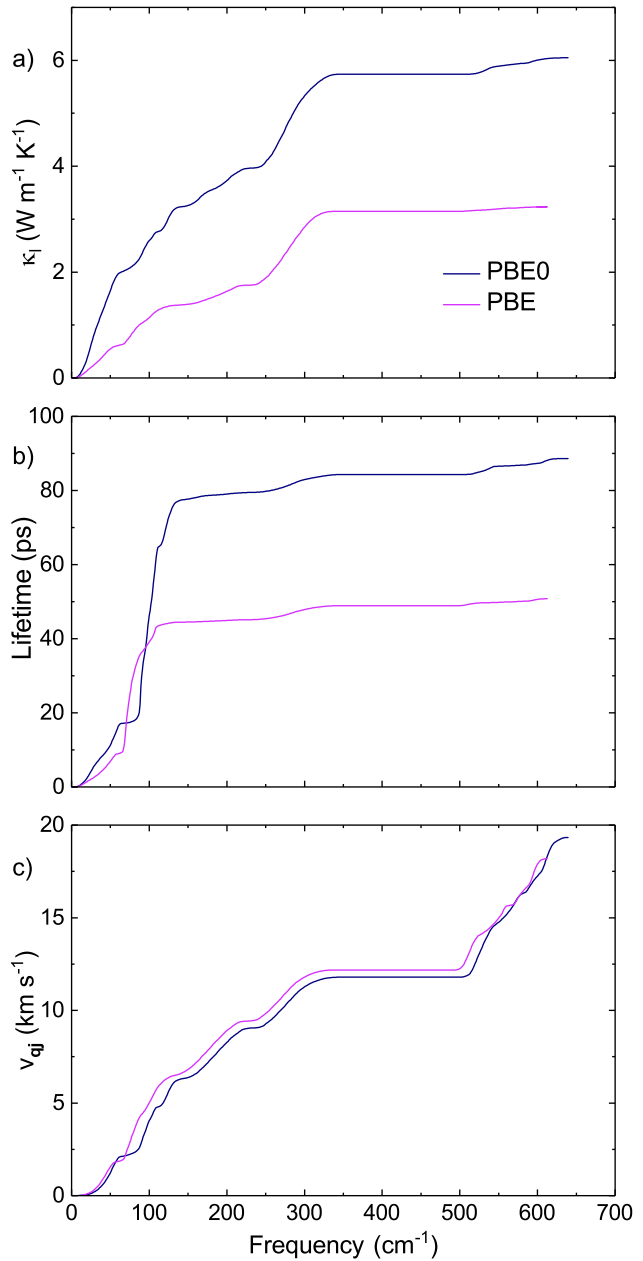}
	\caption{\label{cumulatives}Cumulative properties with respect to harmonic frequencies at 300 K for PBE and PBE0: a) lattice thermal conductivity, b) phonon lifetime and c) group velocity.}
\end{figure}

A more detailed analysis can be made by investigating the different properties through which $\kappa_l$ is calculated. One way is to plot the cumulative properties with respect to phonon frequency as in Figure \ref{cumulatives}. As seen in subplot b), for phonons with $\omega_{\textbf{q}j}$ $\leq$ 250 cm$^{-1}$, the phonon lifetimes obtained with PBE0 are almost twice as large as PBE lifetimes. The difference in lifetimes is even larger in the region below 50 cm$^{-1}$, which explains the much larger cumulative $\kappa_l$ of PBE0 in that regime compared to higher frequencies where the differences in $\tau_{\textbf{q}j}$ between functionals PBE and PBE0 are smaller. Despite the much higher phonon density in frequencies from around 60 to 150 cm$^{-1}$, contribution to thermal conductivity is only half of that from the TA modes below 50 cm$^{-1}$. Group velocities play a bigger role in this regime since even though lifetimes are on the same scale as for the TA modes, given the number of modes in the frequency range, the smaller group velocity which is squared in the formula of $\kappa_l$ dominates the modes effect on thermal conductivity. The phonon modes showing large dispersion between 150 and 350 cm$^{-1}$ make up about one third of the total lattice thermal conductivity despite their rather short lifetimes. Optical modes above 500 cm$^{-1}$ clearly carry less heat than the modes described above. With PBE, the amount of heat carried by the optical modes is almost negligible due to short lifetimes.
	
Another key feature to note in the cumulative properties is that acoustic phonons carry a relatively small portion of the heat in \ce{Cu2O}. For comparison, in the case of \ce{WS2} and \ce{MoS2}, practically all heat is carried by phonons below 150 cm$^{-1}$. This region also includes some optical modes, but still it is safe to assume that the heat carried by acoustic phonons is much greater than in \ce{Cu2O}. Generally, lifetimes in \ce{Cu2O} are relatively short compared to other recently studied semiconductors, such as Si, Si clathrate frameworks, or the aforementioned sulfides.\cite{Lindroth2016,Harkonen2016,Harkonen2016a} Lifetimes of different modes are in the order of picoseconds (see Supplemental Material, Figures S8 and S9) where as for \ce{WS2} and Si, the lifetimes of some modes are already in nanoseconds. The lifetime of a phonon mode is strongly affected by the number of available phonon-phonon scattering pathways, Eq. \ref{Gamma}. In \ce{Cu2O}, the relatively low-lying optical modes around 100 cm$^{-1}$ apparently facilitate the phonon scattering of the acoustic modes. These optical modes are even closer to the TA modes in the case of PBE, which underestimates their energy in comparison to the experiment, increasing the number of scattering pathways, and thus decreasing the lattice thermal conductivity due to shorter phonon lifetimes. This is a somewhat similar effect what Lindroth and Erhart showed for \ce{WS2} by artificially reducing the energy gap between acoustic and optical phonon modes. As optical modes were brought closer in energy towards the heat carrying low-frequency phonons, lifetimes became shorter.

To further investigate the effect, we plotted the imaginary parts of the self-energy, joint density of states (JDOS) and the weighted joint density of states (w-JDOS) with respect to frequency at the R-point (\textbf{q} = (0.5, 0.5, 0.5)) of the first Brillouin zone. We checked the corresponding values also at ten points from $\Gamma$ to the three directions [100], [110] and [111], and the appearances of the plots were similar, so the point R serves as a representative point in the reciprocal space. This gives us explicit information on the number of different pathways for phonon scattering through summing the three-phonon events that do not violate the conservation of energy and momentum. The JDOS is split in two types. Class 1 events represent the different collision processes, effectively the delta functions on the second row of Eq. \ref{Gamma}, and class 2 events represent the decay processes, effectively the delta functions in the first row of Eq. \ref{Gamma}. When the effects of temperature, meaning the actual occupations of different modes are taken into account, we sum also the phonon distribution prefactors of Eq. \ref{Gamma} and this constitutes the w-JDOS.

\begin{figure}
	\includegraphics{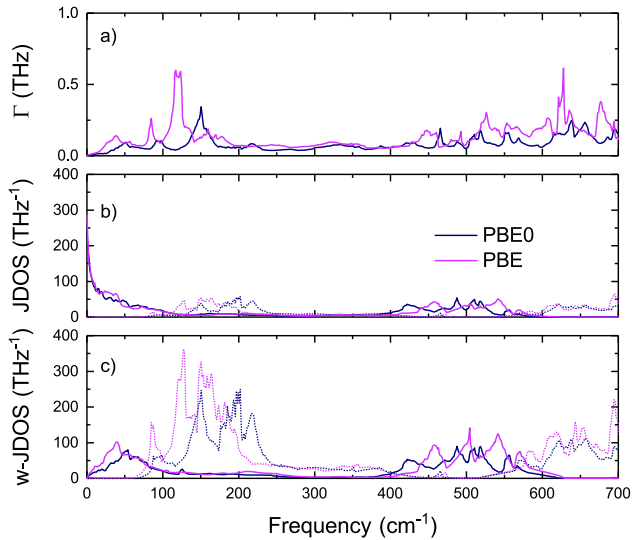}
	\caption{\label{JDOS}The a) imaginary part of the self-energy, b) joint density of states (JDOS) and c) weighted joint density of states(w-JDOS) of \ce{Cu2O} calculated at the R-point in the first Brillouin zone (\textbf{q}= (0.5, 0.5, 0.5)) for both functionals PBE and PBE0. Full lines mark class 1 events and dashed lines class 2 events (see text). $\Gamma$ and w-JDOS are calculated at 300 K. We show the plot only up to 700 cm$^{-1}$, although $\Gamma$ can be nonzero for up to two times the maximum phonon frequency.}
\end{figure}

Figures \ref{JDOS}a and \ref{cumulatives}b show to some degree the same result in reverse. As the imaginary part of the self-energy is inversely proportional to the lifetime, the two plots follow each other in a reciprocal fashion when the two functionals are compared. Subplot \ref{JDOS}b is as expected, there are no decay events until frequencies where there are optical modes. Between 100 and 200 cm$^{-1}$ there are some minor differences, mainly shifts in frequency, but based on the JDOS alone, there wouldn't seem to be much difference in the three-phonon scattering events between the two functionals. Major differences with PBE and PBE0 arise in the w-JDOS. Similarly as in the JDOS, the peaks appear in lower frequencies for PBE because of the general softening of the modes, but here also the magnitudes differ significantly. Since the calculated w-JDOS shows such clear differences between PBE and PBE0, it could give useful information on the performance of difference functionals for calculating $\kappa_l$ already based on phonon harmonic properties. However, only one representative point in the Brillouin zone has been discussed here and in any case, to make a real comparison, one needs to include the anharmonicity in order to see the effect of the interaction strength that is not included in the w-JDOS.

\section{Conclusions}

We have carried out the first \textit{ab initio} DFT study on the lattice thermal conductivity of a semiconductor material using a full hybrid density functional method. In the case of \ce{Cu2O}, the hybrid DFT-PBE0 method outperforms its GGA counterpart DFT-PBE both in the case of lattice thermal conductivity and for various other properties studied here. The largest difference is seen for the lattice thermal conductivity, where PBE0 overestimates $\kappa_l$ by only 5\% in comparison to experiment and PBE underestimates it by over 40\%. Other quantities, such as electronic band gap, phonon frequencies at $\Gamma$-point, and mode-Grüneisen parameters are also closer to the experiment when calculated with PBE0. Overall, hybrid density functional calculations with Gaussian-type local basis sets provide a rather cost-efficient way to investigate the phonon properties of the challenging 3\textit{d} transition metal oxides without any empirical corrections. Work still remains to be done to take into account factors such as point defect scattering and anharmonic frequency shifts.

\begin{acknowledgments}
	We thank Dr. Klaus-Peter Bohnen and Dr. Rolf Heid from Karlsruhe Institute of Technology for providing their inelastic neutron scattering data. The work has been funded by the Academy of Finland (Strategic Research Council, CloseLoop consortium, grant 303452). Computational resources were provided by CSC, the Finnish IT Center for Science.
\end{acknowledgments}

\end{document}